# Patient-specific Finite Element Modeling of Aneurysmal dilatation after chronic type B aortic dissection

**Shaojie Zhang[a], Joan D. Laubrie[a], S. Jamaleddin Mousavi[a], Stéphane Avril[a] and Sabrina Ben Ahmed[b]**

[a] Mines Saint-Etienne, Univ Lyon, Univ Jean Monnet, INSERM, U1059 Sainbiose, Centre CIS, F-42023 Saint-Etienne, France.

[b] Univ Jean Monnet, INSERM, U1059 Sainbiose and University Hospital of Saint-Etienne, F-42000 Saint-Etienne,

**Abstract** **Progressive aneurysmal dilatation is a well-recognized complication in patients with chronic type B aortic dissection (cTBAD), which may lead to a delayed rupture and create a life-threatening condition. However, our understanding of such aortic expansion in cTBAD remains weak. In the present paper, we propose to use numerical simulations to study the role of growth and remodeling (G&R) in aneurysmal dilatation after cTBAD. We set up a 3D finite-element model of G&R for aortic dissection within an open-source code. Constitutive equations, momentum balance equations, and equations related to the mechanobiology of the artery were formulated based on the homogenized constrained mixture theory. The model was first applied to idealized aortic geometries with cylindrical and toric shapes to demonstrate its feasibility and efficiency. The model was then applied to a patient-specific aortic segment to show its potential in more relevant and complex patient-specific clinical applications. It was found that the G&R tends to naturally trigger the aneurysmal dilatation after dissection, in order to restore its tensional equilibrium. Our results indicated that the value of the gain parameter, related to collagen G&R, plays an important role in the stability of aortic expansion after cTBAD. A small gain parameter will induce an excessive aneurysmal degeneration whilst a large gain parameter helps to recover a stabilized state of the artery after dissection. Finally, it was found that other mechanobiology-related parameters, such as the circumferential length of the dissection, as well as the pressure in the false lumen, may also be determinant for the stability of aneurysmal dilatation after cTBAD. Both a wide tear and an elevated false lumen pressure favor an unstable development of aortic expansion after cTBAD. As future work, the present model will be validated through predictions of aneurysmal dilatation in patient-specific clinical cases, in comparison with datasets followed over a significant period of time.**

# I. INTRODUCTION

Chronic type B aortic dissection (cTBAD) is defined when a tear originates in the descending aorta and remains 3 months after its onset [1]. Patients with uncomplicated cTBAD are preferentially treated medically with periodic clinical and imaging surveillance, regarding the acceptable survival rate generally observed in a short-term follow-up [2], [3]. However, the long-term outcome of such conservative management remains questionable mainly due to the progressive aneurysmal dilatation [4]. Invasive surgical interventions, such as endovascular repair or open surgery are then needed [5]. Up to now, little is known about the aneurysmal dilatation after cTBAD, either it is stable with a moderated progression rate or there is an excessive aneurysmal degeneration. It is yet of crucial importance for surgeons to be able to assess the risk of aortic expansions in patients with early-stage cTBAD to choose the optimal treatment approach. Patients identified at high risk for aortic enlargement may therefore benefit from early surgical interventions and reduce mortality from delayed aneurysm ruptures.

Published studies on this topic remain scarce. It has been widely accepted that the presence of an excessive aortic diameter, typically greater than 40 mm, and a patent false lumen are two high-risk factors for late aneurysm development after cTBAD [6]-[8]. Besides, older age and elevated mean blood pressure were also found to promote aneurysmal degeneration in cTBAD [6]. Tsai et al. reported that the size, the number, as well as location of tears have significant impacts on the pressure in the false lumen, and therefore influencing the false lumen expansion [9]. Recently, Trimarchi et al. revealed that there are many other factors that may affect aneurysmal dilatation after cTBAD, including demographic, clinical, pharmacologic, and radiologic variables, such as connective tissues disorders, gender, the presence of thrombus in the false lumen, etc [10]. However, all the above researches were based on observational studies or clinical trials with data collected over a long follow-up period.

Considering the recent advances in computational mechanics of arteries [11], [12] and more specifically the growth and remodeling (G&R) models [13]-[17], numerical models can be an interesting alternative option for studying these influencing factors. However, to the author's best knowledge, G&R after cTBAD has never been modeled so far. There is still an important potential for G&R models to understand vascular adaptation in chronic type B aortic dissection, where the patient can undergo a long-term process of G&R after breaking the initial mechanical equilibrium due to tear opening.

In this specific context, we developed a 3D finite-element model of vascular adaptation to study the aneurysmal dilatation after cTBAD, within an open-source code written in python and C++ [18], [19]. The G&R model of the arterial wall is based on the homogenized constrained mixture theory (CMT) and the aortic dissection is modeled through an original two-continuum arterial wall concept. We also performed a sensitivity analysis to evaluate the influence of several selected mechanobiological parameters on the aneurysmal dilatation after cTBAD.

Details of the model are given in this book chapter, by first introducing the mathematical framework of the CMT method for G&R with respect to cTBAD, then describing the two-continuum aortic dissection model, and finally showing potentials of the model, from a simple validation test case to academic applications with idealized geometries, until a more relevant patient-specific application.

# II. Material and Methods

## A. Constitutive and balance equations

The CMT was first proposed by Humphrey and Rajagopal as a hybrid method to describe mechano-regulated G&R of arteries [20]. It was then largely used for modeling aneurysm formation [13]-[17]. In this work, we employ the homogenized CMT [17] to model arterial G&R after cTBAD. Basic equations formulated under the homogenized CMT framework are briefly introduced in this section. Readers can refer to reference publications for more detailed mathematical formulations and their interpretations [15], [17], [20].

First, we assume that the arterial wall can be modeled as an homogenized mixture made up by a matrix containing a network of elastic fibers, passive reinforcements represented by 4 collagen fiber families (respectively oriented in circumferential, axial and diagonal (+/- 45°) direction) and active reinforcements accounting for the contractility of smooth muscle cells (SMCs) in the circumferential direction. Let $\Omega_R \subset \mathbb{R}^3$ and $\Omega_t \subset \mathbb{R}^3$ denote, respectively, the initial traction-free reference configuration at time $t = 0$ and current deformed configuration at time $t > 0$ of the arterial wall. According to homogenized CMT, we assume that all constituents in the arterial wall deform together with a same deformation gradient $\mathbf{F}$:

$$\mathbf{F} = \frac{\partial \boldsymbol{x}}{\partial \boldsymbol{X}} \tag{1}$$

where $\boldsymbol{X}$ represents a material point in $\Omega_R$ and $\boldsymbol{x}$ represents the associated spatial point in $\Omega_t$. Moreover, based on the theory of Rodriguez and Hoger [21], this deformation gradient tensor $\mathbf{F}$ can be split into an elastic part and an inelastic part for each constituent $i \in [e, c_j, m]$, such as

$$\mathbf{F} = \mathbf{F}_{el}^{i}\mathbf{F}_{gr}^{i} \tag{2}$$

where $e, c_j, m$ represents respectively the elastic matrix, the $j^{th}$ collagen fiber family and smooth muscle cells. More precisely, $\mathbf{F}_{el}^{i}$ represents the elastic deformation tensor related to stresses that balance external mechanical loads over the arterial wall, while $\mathbf{F}_{gr}^{i}$ represents the inelastic deformation tensor related to G&R, i.e. related to the continuous mass turnover of each constituent. Besides, we assume that G&R is a fully stress-mediated process. Other non-mechanical effects related to the mass turnover, such as immune-mediated chemical remodeling, damage, or mechanical fatigue, are neglected in this work. Therefore, the temporal homogenized mass deposition or degradation rate of each constituent can be expressed as

$$\dot{\rho}_R^i = \rho_R^i k_\sigma^i \frac{\sigma^i - \sigma_h^i}{\sigma_h^i} \tag{3}$$

where $\rho_R^i$ is the reference mass density of constituent $i$, related to the reference configuration of the arterial wall. The right-hand side term of Equation 3 describes the mass turnover due to the stress difference between the current stress $\sigma^i$ and the homeostatic stress $\sigma_h^i$, where $k_\sigma^i$ is a regularization parameter (named gain parameter) with respect to each constituent.

The homogenized CMT consists in the decomposition of the inelastic deformation gradient $\mathbf{F}_{gr}^i$ through two sub-gradient tensors

$$\mathbf{F}_{gr}^i = \mathbf{F}_g^i \mathbf{F}_r^i \tag{4}$$

where $\mathbf{F}_g^i$ is the growth-related tensor describing volume changes to due mass turnover, and $\mathbf{F}_r^i$ is the remodeling-related tensor describing how the prestretch of each constituent is updated through continuous extant mass degradation and new mass production. As suggested by Braeu et al. [15], we assume that the growth deformation is the same for all constituents in the arterial wall, such as

$$\mathbf{F}_g^i = \mathbf{F}_g = \mathbf{I} + \frac{\rho_R}{\rho_{R0}} \boldsymbol{a}_0^\perp \otimes \boldsymbol{a}_0^\perp - \boldsymbol{a}_0^\perp \otimes \boldsymbol{a}_0^\perp \tag{5}$$

where $\rho_R$ is the current reference mass density, $\rho_{R0}$ is the initial reference mass density (at time $t = 0$), $\mathbf{I}$ is the second order identity and $\boldsymbol{a}_0^\perp$ the growth direction [17]. The remodeling process of elastin can be generally neglected ($\mathbf{F}_r^e = \mathbf{I}$) considering its slow mass degradation rate (typically several decades for elastin half-life time). We assume that the remodeling process of collagen fibers and SMCs occurs at a constant volume and along the fiber direction, which can be expressed as [22]

$$\mathbf{F}_r^i = \lambda_r^i \boldsymbol{a}_0^i \otimes \boldsymbol{a}_0^i + \frac{1}{\sqrt{\lambda_r^i}} \left(\mathbf{I} - \boldsymbol{a}_0^i \otimes \boldsymbol{a}_0^i\right) \tag{6}$$

where $\lambda_r^i$ is the respective remodeling stretch of fiber $i \in [c_j, m]$ along its fiber direction $\boldsymbol{a}_0^i$ with its time evolution $\dot{\lambda}_r^i$ given by [13]

$$\dot{\lambda}_r^i = \left(\frac{\dot{\rho}_R^i}{\rho_R^i} + \frac{1}{T^i}\right) \frac{\lambda^i}{\left(\lambda_{el}^i\right)^2} \left(\frac{\partial \sigma^i}{\partial \lambda_{el}^i}\right)^{-1} \times \left(\sigma^i - \sigma_h^i\right) \tag{7}$$

where $T^i$ is the average mass turnover time during which old mass increment is degraded and replaced by a new mass increment. $\lambda_{el}^i$ is the elastic stretch of fiber $i$ defined as $\lambda_{el}^i = \sqrt{\left(\mathbf{F}_{el}^i\right)^t \mathbf{F}_{el}^i : \boldsymbol{a}_0^i \otimes \boldsymbol{a}_0^i}$ and $\lambda^i$ is the total stretch of fiber $i$ defined as $\lambda^i = \sqrt{(\mathbf{F})^t \mathbf{F} : \boldsymbol{a}_0^i \otimes \boldsymbol{a}_0^i}$

Finally, considering that the homeostatic configuration of the arterial wall is achieved at time $t = t_0$ and defining the initial traction-free geometry of the arterial wall at time $t = 0$ as the same geometry as its homeostatic configuration, the initial elastic prestretch of each constituent $\mathbf{G}_h^i$ corresponding to the homeostatic configuration at time $t_0$ can simply satisfy

$$\mathbf{F}_r^i(t_0) = \left(\mathbf{G}_h^i\right)^{-1} \tag{8}$$

due to the fact that $\mathbf{I} = \mathbf{F}(t_0) = \mathbf{F}_e^i(t_0)\mathbf{F}_g^i(t_0)\mathbf{F}_r^i(t_0) = \mathbf{G}_h^i\mathbf{F}_g^i(t_0)\mathbf{F}_r^i(t_0) = \mathbf{G}_h^i\mathbf{F}_r^i(t_0)$. The detailed expressions of $\mathbf{G}_h^i$ are hereby given for each constituent $i \in [e, c_j, m]$ , with respect to a cylindrical coordinate system

$$\mathbf{G}_h^e = \mathbf{diag}\left(\frac{1}{\lambda_\theta^e\lambda_z^e},\ \lambda_\theta^e,\ \lambda_z^e\right) \tag{9}$$

$$\boldsymbol{G}_h^{c_i} = \lambda_h^{c_i}\boldsymbol{a}_0^{c_i} \otimes \boldsymbol{a}_0^{c_i} + \frac{1}{\sqrt{\lambda_h^{c_i}}}\left(\boldsymbol{I} - \boldsymbol{a}_0^{c_i} \otimes \boldsymbol{a}_0^{c_i}\right) \tag{10}$$

$$\boldsymbol{G}_h^{m} = \lambda_h^{m}\boldsymbol{a}_0^{m} \otimes \boldsymbol{a}_0^{m} + \frac{1}{\sqrt{\lambda_h^{m}}}\left(\boldsymbol{I} - \boldsymbol{a}_0^{m} \otimes \boldsymbol{a}_0^{m}\right) \tag{11}$$

where $\lambda_\theta^e$ and $\lambda_z^e$ are the initial deposition stretches of elastin, respectively in the circumferential and axial direction, uniform over the whole arterial wall. $\lambda_h^{c_i}$ and $\lambda_h^m$ are respectively the initial deposition stretches of collagen fibers (same deposition stretch for all four directions) and SMCs.

Based on CMT, the strain energy density function of the arterial wall can be given by

$$\Psi = \rho_R^e W^e + \sum_{j=1}^{4} \rho_R^{c_j} W^{c_j} + \rho_R^m W^m \tag{12}$$

where $\rho_R^e$, $\rho_R^{c_j}$ and $\rho_R^m$ are respectively the reference mass densities of the elastic matrix, of the $j^{th}$ collagen fiber family and of SMCs, and $W^e, W^{c_j}$ and $W^m$ are the specific strain energy density functions with respect to each constituent. Moreover, the strain energy density function $W^i$ of each constituent $i \in [e, c_j, m]$, can be expressed as a function of its elastic deformation tensor $\mathbf{F}_{el}^i$, or equivalently, its elastic right Cauchy-Green tensor $\mathbf{C}_{el}^i$, defined as $\mathbf{C}_{el}^i = \left(\mathbf{F}_{el}^i\right)^t\mathbf{F}_{el}^i$. In the present work, the elastic matrix is considered as a quasi-incompressible Neo-Hookean hyperelastic material with its specific strain energy density function $W^e$ given by

$$W^e = \frac{\mu^e}{2}(\mathrm{tr}(\bar{\mathbf{C}}_{el}^e) - 3) + \kappa(|\mathbf{F}_{el}^e| - 1)^2 \tag{13}$$

where $\mu^e$ is a material parameter characterizing the shear stiffness of elastin and $\kappa$ is an arbitrary but sufficiently high penalty parameter ensuring quasi incompressibility. $\bar{\mathbf{C}}_{el}^e$ is the isochoric elastic right Cauchy-Green tensor of elastin, defined as $\bar{\mathbf{C}}_{el}^e = (\bar{\mathbf{F}}_{el}^e)^t\bar{\mathbf{F}}_{el}^e$ and $\bar{\mathbf{F}}_{el}^e = \mathbf{F}_{el}^e/|\mathbf{F}_{el}^e|^{1/3}$. The specific strain energy density function of collagen fiber families is modeled by an anisotropic Fung-type exponential function,

$$W^{c_j} = \frac{k_1^{c_j}}{2k_2^{c_j}}\left(e^{k_2^{c_j}\left(I_{4_{el}}^{c_j}-1\right)^2} - 1\right) \tag{14}$$

We also use the same anisotropic Fung-type exponential function to model the passive behavior of SMCs [15], while an additional term is added for the active tone contribution such as,

$$W^m = \frac{k_1^m}{2k_2^m}\left(\boldsymbol{e}^{k_2^m\left(I_{4_{el}}^m-1\right)^2} - 1\right) + \frac{\sigma_{max}}{\rho_{R0}}\left(\lambda_{act} + \frac{1}{3}\frac{(\lambda_{max}^m - \lambda_{act})^3}{(\lambda_{max}^m - \lambda_0^m)^2}\right) \tag{15}$$

where $k_1^{c_j}$ and $k_1^m$ are stress-like material parameters, and $k_2^{c_j}$ and $k_2^m$ are dimensionless material parameters. $I_{4_{el}}^{c_j}$ and $I_{4_{el}}^m$ are pseudo-invariants, which are additional invariants defined in case of anisotropic materials such as $\lambda_{el}^i = \mathbf{C}_{el}^i : \boldsymbol{a}_0^i \otimes \boldsymbol{a}_0^i$ with $i \in [c_j, m]$ [23]. $\lambda_{act}$ is the active stretch in the circumferential direction, $\sigma_{max}$ is the maximum active Cauchy stress, $\rho_{R0}$ is the reference total mass density of the arterial wall at time $t = 0$, and $\lambda_{max}^m$ and $\lambda_0^m$ are the active stretches respectively at maximum and zero active stress for SMCs.

The second Piola-Kirchhoff stress tensor $\mathbf{S}$ and the fourth order elasticity tensor of the arterial wall $\mathbb{C}$ are then deduced by performing the first and second derivatives of the strain energy function $\Psi$ with respect to the total Green-Lagrange strain $\mathbf{E}$

$$\mathbf{S} = \frac{\partial \Psi}{\partial \mathbf{E}} = \varphi^e \mathbf{S}^e + \sum_i \varphi^{c_j} \mathbf{S}^{c_j} + \varphi^m \mathbf{S}^m \tag{16}$$

$$\mathbb{C} = \frac{\partial^2 \mathbf{S}}{\partial \mathbf{E} \partial \mathbf{E}} = \varphi^e \mathbb{C}^e + \sum_i \varphi^{c_j} \mathbb{C}^{c_j} + \varphi^m \mathbb{C}^m \tag{17}$$

where $\varphi^i$, $\mathbf{S}^i$ and $\mathbb{C}^i$ are the mass fraction, second Piola-Kirchhoff stress and forth order elasticity tensor with respect to each constituent $i \in [e, c_j, m]$ in the arterial wall, defined as

$$\varphi^i = \frac{\rho_R^i}{\rho_R} \tag{18}$$

$$\mathbf{S}^i = \rho_R \frac{\partial W^i}{\partial \mathbf{E}} \tag{19}$$

$$\mathbb{C}^i = \rho_R \frac{\partial^2 W^i}{\partial \mathbf{E} \partial \mathbf{E}} \tag{20}$$

with $\rho_R = \rho_R^e + \rho_R^{c_j} + \rho_R^m$ the reference total mass density of the arterial wall. Finally, assuming that the G&R occurs at a slow time scale and can be considered as a quasi-static process, the dynamics effects such as inertia or viscoelasticity of the arterial wall can be neglected. Therefore, the momentum balance equations of the arterial wall can be simply written as

$$\nabla \cdot \boldsymbol{\sigma} + \rho \boldsymbol{b} = \boldsymbol{0} \tag{21}$$

$\rho$ is the spatial density of the arterial wall, related to its reference density $\rho_R$, as $\rho = \rho_R/|\mathbf{F}|$, $\boldsymbol{b}$ is the body force vector given in the spatial configuration, $\boldsymbol{\sigma}$ is the Cauchy stress derived from the previous second Piola-Kirchoff stress as

$$\boldsymbol{\sigma} = \frac{\mathbf{1}}{|\mathbf{F}|}\mathbf{FSF}^t \tag{22}$$

The boundary conditions applied on the arterial wall can be Dirichlet boundary conditions, assigning the predefined displacement field over the mesh nodes or Robin boundary conditions, which are applied over the mesh surface, such as

$$\boldsymbol{\sigma} \cdot \boldsymbol{n} = P_{TL}\boldsymbol{n} + \boldsymbol{p}_{dissection} + \boldsymbol{F}_{spring} \tag{23}$$

where $P_{TL}$ denotes the true luminal pressure of the artery due to blood flow, applied on the inner surface of the media layer. $\boldsymbol{n}$ is the outward pointing unit vector normal to the arterial inner media surface. $\boldsymbol{p}_{dissection}$ is the pressure in the false lumen after cTBAD. $\boldsymbol{F}_{spring}$ is an additional spring-based elastic force, related to the two-continnumm arterial wall concept proposed in this work. Details of the two-continuum arterial wall concept, as well as, $\boldsymbol{F}_{spring}$, $\boldsymbol{p}_{dissection}$, will be given in the next section.

### B. Dissection model

In this section, we will firstly present the two-continuum arterial wall concept, dedicated to the modeling of G&R in the case of cTBAD. It is worth noting that the mechanism of the initial tear formation or the subsequent tear progression is currently not of primary interest in this paper. Under this specific context, we propose to model the initial healthy arterial wall without dissection with two continuum bodies. As shown in Figure 1a, the arterial wall is made up of two layers, respectively the inner media layer and the outer adventitia layer. The two layers are perfectly connected by high stiffness elastic springs. More precisely, each spring connects two adjacent mesh surfaces $S^a$ and $S^m$, respectively at the inner surface of the adventitia and the outer surface of media. The force applied on each surface is computed on the relative displacement of the two connecting surfaces, as given in Equation (24)

$$\boldsymbol{F}_{dissection} = \begin{cases} -k_{spring}(d_{S^a} - d_{S^m})\boldsymbol{n}_{S^a}, & adventitia \\ \ \ k_{spring}(d_{S^a} - d_{S^m})\boldsymbol{n}_{S^m}, & media \end{cases} \tag{24}$$

where $k_{spring}$ is the stiffness of the interfacial spring. $d_{S^a}$ and $d_{S^m}$ the nodal-averaged displacement of the mesh surface located respectively in the inner adventitia and outer media surfaces. $\boldsymbol{n}_{S^a}$ and $\boldsymbol{n}_{S^m}$ are the outward pointing unit vectors normal to the respective mesh. It is worth noting that displacement of one layer is transmitted to the other layer with low distortions due to the high stiffness of the springs. Finally, we assume that this initial configuration of the arterial wall is at its homeostatic state with a reference luminal pressure $P_{TL}$. To validate this new concept, a simple validation test case has been proposed in this work, in comparison with the conventional arterial wall model with a single continuum body [13], [15], [16], [17].

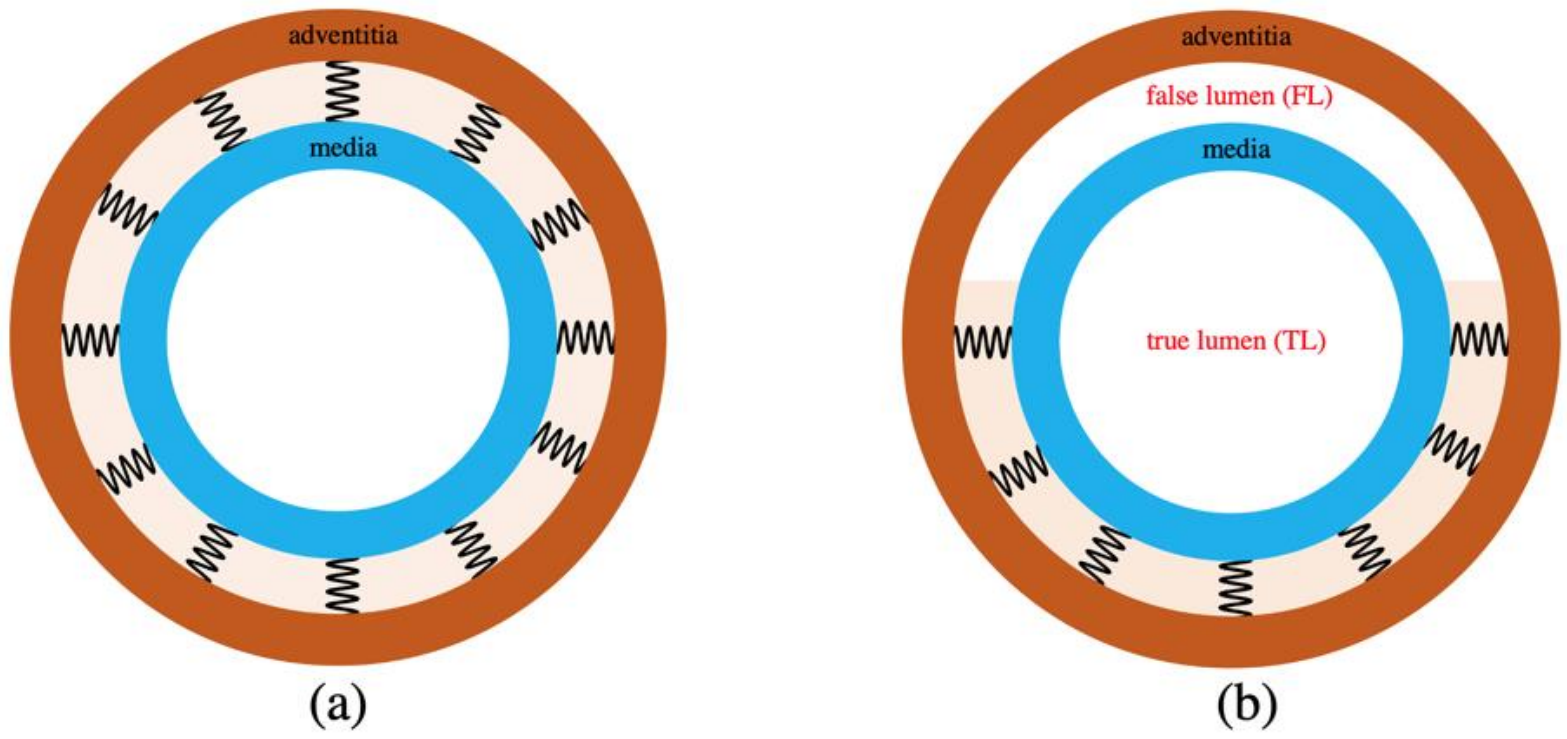

Figure 1 - (a) Illustration of the two-continuum arterial wall model, with the adventitia layer and the media layer connected by high stiffness elastic springs, representing a healthy arterial wall under its homeostatic state. (b) Initial configuration of the arterial wall after cTBAD. The false lumen is created by breaking interfacial elastic springs in a selected region between the media and adventitia layers.

Based on this new concept of a two-continuum arterial wall, the initial configuration of cTBAD with false lumen can be obtained by simply vanishing interfacial springs. As shown in Figure 1b, the false lumen (FL), as well as the free surfaces (*i.e.* inner side of adventitia layer and outer side of media layer), is created by breaking springs in a selected region where tears are assumed to be present. The force induced by the pressure in the false lumen is then applied on each mesh surface, $S^{dissection}$, of the newly created free surfaces

$$\boldsymbol{p}_{dissection} = P_{FL}\boldsymbol{n}_{S^{dissection}} \tag{25}$$

where $P_{FL}$ is the constant pressure in the false lumen and $\boldsymbol{n}_{S^{dissection}}$ is the outward pointing unit vector normal to the free mesh surfaces in the false lumen. Finally, it is worth noting that the presence of such pressurized false lumen will break the mechanical equilibrium of the arterial initial homeostatic state, and therefore trigging the G&R of the arterial wall over a long period of time in case of cTBAD, until the achievement of a new preferred mechanical state or eventually an excessive aneurysmal degeneration.

### C. Finite element implementation

The proposed model was implemented within an open-source finite element code, written in Python and C++ [18], [19]. Three different steps, are defined in the model:

- **1st step**: Computation of the healthy arterial wall at homeostasis. The initial arterial wall is loaded with a constant luminal pressure, $P_{TL}$, on its entire inner surface of media.
- **2nd step**: Opening of the arterial wall. Interfacial springs are removed in a selected region between the adventitia layer and media layer, creating the initial dissection tear of cTBAD. The same luminal pressure in the true lumen, $P_{TL}$, is maintained as in the previous step, applied on the entire inner surface of the media. In the meantime, the false lumen will be loaded with a constant pressure, $P_{FL}$, applied on the newly created surfaces related to the tear-open region.
- **3rd step**: Adaptation of the arterial wall with G&R after cTBAD, *i.e.*, after the creation of a pressurized false lumen.

Note that there is only one time increment in the first two steps while the third step is composed of several time increments to obtain relevant results of a long-term arterial wall adaptation after cTBAD. For each time increment, the same set of momentum balance and constitutive equations is solved with the Newton-Raphson method. Finally, at the end of each time increment, we obtain the displacement field and the associated stress-strain information on each mesh node of the arterial wall.

## III. Numerical applications

In order to show the potentials of the present dissection model, four different simulations were performed in this paper:

- **Validation test case**: It consists of a simple test case to validate the spring-connected two-continuum arterial wall concept proposed in this work. The validation was achieved through the comparison with reference results in the literature [13], [15], [16], [17] where the conventional single continuum arterial wall model was employed to simulate the aneurysm formation in response to external stimuli.
- **Application to a cylindrical artery**: The second application is to study the G&R after cTBAD in the case of an idealized cylindrical artery.
- **Application to a toric artery**: In this third application, the dissection model is applied to an idealized toric artery.
- **Application to a patient-specific artery**: In this last application, we demonstrate the feasibility of our dissection model for further more complex and relevant clinical patient-specific applications, by modeling the G&R on a dissected patient-specific human descending aortic segment.

## A. Validation test case

An idealized two-layered cylindrical artery is considered. The geometry was the same as that has been used in the work of Braeu et al. [15], with an inner arterial radius of 10 mm and a constant arterial thickness of 1.41 mm. Besides, we assume that each layer of the arterial wall, *i.e.* adventitia layer and media layer, has the same thickness of 0.705 mm. Moreover, it should be mentioned that a constant gap of 0.01 mm is defined between the two layers, allowing the presence of interfacial connecting springs with respect to the two-continuum arterial wall concept. The mesh was hexahedral and composed of 6 × 40 × 25 elements (thickness × circumferential × length). Finally, the whole geometry is assumed to be at a homeostatic state, related to a reference luminal pressure of 13.3 kPa.

Following Braeu et al. [15], we apply a sudden degradation of the elastic matrix such as

$$\dot{D}^e = -\frac{\rho_R^e}{T^e} - \frac{D_{dam}}{t_{dam}} \rho_{R0}^e e^{-0.5\left(\frac{z}{L_{dam}}\right)^2 - \frac{t}{t_{dam}}} \tag{26}$$

where $L_{dam}$ and $t_{dam}$ are respectively the spatial and the temporal damage spread parameter, $D_{dam}$ is the maximum damage, $\rho_{R0}^e$ is the initial reference mass density of elastin, $z$ is the axial position of the cylinder. Noting that due to the symmetry of the problem, only half of the cylinder is modeled in this simulation with $z$ varying from 0 mm to 90 mm. The first term at the right-hand side (RHS) of Equation 26 describes the natural elastin degradation by aging effect. The second term of the RHS is related to sudden external stimuli, causing a maximum elastin degradation at the center of the cylinder, i.e. $z = 0$ mm. Material properties used in this validation test case are taken from the 3D model of Braeu et al. [15] and are summarized in Table 1 together with other simulation parameters. The simulation results obtained in this simulation are compared to the reference results in the literature [13], by using three different values of collagen gain parameter $k_\sigma^{c_i}$.

Table 1. Material properties used in the validation test case.

| Symbol | Value | Unit |
|---|---|---|
| $\mu^e$ | 72 | $\mathrm{J \cdot kg^{-1}}$ |
| $\kappa$ | 720 | $\mathrm{J \cdot kg^{-1}}$ |
| $k_1^{c_j}$ | 568 | $\mathrm{J \cdot kg^{-1}}$ |
| $k_2^{c_j}$ | 11.2 | – |
| $k_1^m$ | 7.6 | $\mathrm{J \cdot kg^{-1}}$ |
| $k_2^m$ | 11.4 | – |
| $\rho_{R0}^e$ | 241.5 | $\mathrm{kg \cdot m^{-3}}$ |
| $\rho_{R0}^{c_1} = \rho_{R0}^{c_2}$ | 65.1 | $\mathrm{kg \cdot m^{-3}}$ |
| $\rho_{R0}^{c_3} = \rho_{R0}^{c_4}$ | 260.4 | $\mathrm{kg \cdot m^{-3}}$ |
| $\rho_{R0}^{m}$ | 157.5 | $\mathrm{kg \cdot m^{-3}}$ |
| $\lambda_z^e$ | 1.25 | – |
| $\lambda_\theta^e$ | 1.34 | – |
| $\lambda_h^{c_j}$ | 1.062 | – |
| $\lambda_h^m$ | 1.1 | – |
| $T^e$ | 101. 16 | $years$ |
| $T^{c_j}$ | 101 | days |
| $T^m$ | 101 | days |
| $L_{dam}$ | 8 | mm |
| $t_{dam}$ | 40 | days |
| $D_{max}$ | 0.5 | – |
| $\sigma_{actmax}$ | 54 | kPa |
| $\lambda_0$ | 0.8 | – |
| $\lambda_m$ | 1.4 | – |
| $k_{spring}$ | 1000 | $\mathrm{kPa \cdot mm^{-1}}$ |

### B. Application to a cylindrical artery

After the validation of the two-continuum arterial wall model, we first apply the G&R after cTBAD in the case of an idealized two-layered cylindrical artery as shown in Figure 2. The same geometry as in the previous validation case was used, except that the length of the artery is reduced from 90 mm to 50 mm. The mesh was hexahedral and composed of 6 × 60 × 20 elements (thickness × circumferential × length). Similarly, we assume that this geometry was related to a homeostatic state under an inner true lumen pressure of 100 mmHg. The initial tear of the dissection is created by breaking springs in regions where $x \leq 10$ and $z \leq 50$, to model a representative initial tear of cTBAD. A pure sliding boundary condition is assigned on the cross-section at two extremities. The outer surface of the adventitia is free. The reference pressure in the false lumen is assumed to be the same as in the true lumen. The same material properties as reported in Table 1 has been used, considering additionally the layered distribution of different material constituents as suggested by Mousavi et al. [16] for human ascending thoracic aorta, i.e., the media has 97% of the total elastin, 15% of the total axial and diagonal collagen fibers, and 100% of the total SMCs, while the adventitia has 3% of the total elastin, 85% of the total axial and diagonal collagen and 100 % of the total circumferential collagen.

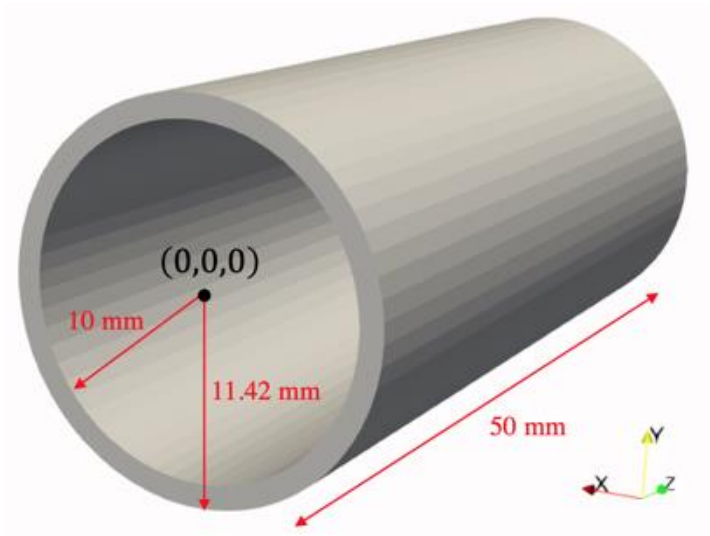

Figure 2 – Schematic representation of the idealized cylindrical artery on which the presented dissection model is applied for modeling G&R after cTBAD.

### C. Application to a toric artery

In order to further verify the applicability of the present dissection model, we employed here an idealized toric geometry, as shown in Figure 3, to simulate the arterial G&R after cTBAD.

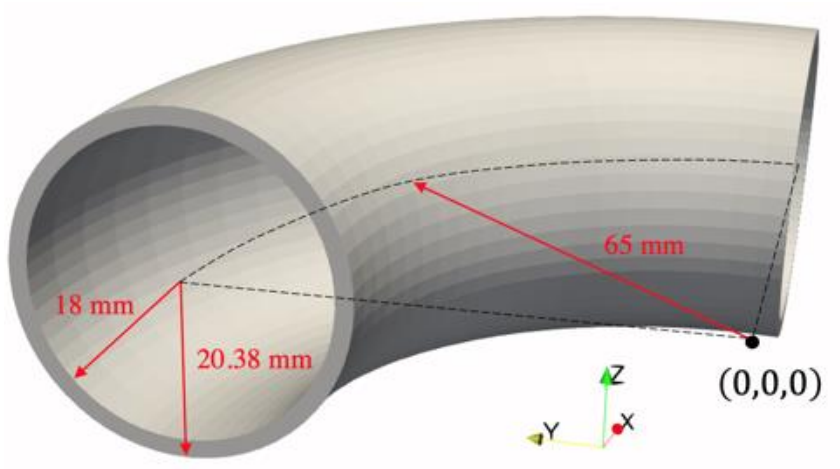

Figure 3 - Schematic representation of the idealized toric artery on which the present dissection model was applied for modeling G&R after cTBAD.

The geometry was a fourth of a torus with an arch radius of 65 mm, similar to the ones already used in the literature [16], [24]. The arterial section is defined with an inner radius of 18 mm and an outer radius of 20.38 mm. The thickness of the artery is 2.38 mm, including a constant gap of 0.01 mm between two equal-thickness adventitia and media layers. The mesh was hexahedral and composed of 6 × 60 × 20 elements (thickness × circumferential × length). Once again, we assume that this geometry was related to a homeostatic state, under a constant inner true lumen pressure of 80 mmHg. Similarly, a pure sliding boundary condition is assigned to cross-sections at the two extremities while the outer surface of the adventitia is let free. Mechanical parameters, as well as simulation parameters, are reported in Table 2, based on Laubrie et al. [25]. The initial tear of the dissection is created by breaking springs in the regions defined by $\sqrt{x^2+y^2} \geq 70$ and $arctan(x/y) \leq 60°$, to model a representative initial tear of cTBAD.

Table 2. Material properties and simulation parameters used for the toric artery simulation.

| Symbol | Value | Unit |
|---|---|---|
| $\mu^e$ | 80 | $\mathrm{J \cdot kg^{-1}}$ |
| $\kappa$ | 800 | $\mathrm{J \cdot kg^{-1}}$ |
| $k_1^{c_j}$ | 292.0 | $\mathrm{J \cdot kg^{-1}}$ |
| $k_2^{c_j}$ | 5.6 | – |
| $k_1^m$ | 13.8 | $\mathrm{J \cdot kg^{-1}}$ |
| $k_2^m$ | 6.0 | – |
| $\rho_{R0}^e$ (media) | 169.0 | $\mathrm{kg \cdot m^{-3}}$ |
| $\rho_{R0}^{c_1} = \rho_{R0}^{c_2}$ (media) | 14.6 | $\mathrm{kg \cdot m^{-3}}$ |
| $\rho_{R0}^{c_3} = \rho_{R0}^{c_4}$ (media) | 58.4 | $\mathrm{kg \cdot m^{-3}}$ |
| $\rho_{R0}^m$ (media) | 735.0 | $\mathrm{kg \cdot m^{-3}}$ |
| $\rho_{R0}^e$ (adventitia) | 565.0 | $\mathrm{kg \cdot m^{-3}}$ |
| $\rho_{R0}^{c_1} = \rho_{R0}^{c_2}$ (adventitia) | 48.5 | $\mathrm{kg \cdot m^{-3}}$ |
| $\rho_{R0}^{c_3} = \rho_{R0}^{c_4}$ (adventitia) | 194.0 | $\mathrm{kg \cdot m^{-3}}$ |
| $\rho_{R0}^m$ (adventitia) | 0.0 | $\mathrm{kg \cdot m^{-3}}$ |
| $\lambda_h^{c_j}$ | 11 | – |
| $\lambda_h^m$ | 1.1 | – |
| $T^e$ | 101. 16 | $y$ears |
| $T^{c_j}$ | 101 | days |
| $T^m$ | 101 | days |
| $\sigma_{actmax}$ | 54 | kPa |
| $\lambda_0$ | 0.8 | – |
| $\lambda_m$ | 1.4 | – |
| $k_{spring}$ | 1000 | $\mathrm{kPa \cdot mm^{-1}}$ |

## D. Application to a patient-specific artery

In this last test case, our dissection model was applied to a patient-specific human descending thoracic aortic segment, as shown in Figure 4a. It was taken from a patient-specific aortic arch geometry, reconstructed from a patient's CT scan [16], as shown in Figure 4b. The exact location of the modeled aortic segment is shown in Figure 4b with the blue mesh. The thickness of the adventitia and media was assumed to be equal. Besides, the two layers were separated with a constant gap of 0.01 mm. The mesh was hexahedral and composed of 6 × 48 × 42 elements (thickness × circumferential × length). Similar to previous test cases, we assumed that this initial geometry was related to a homeostatic state, with an inner true lumen pressure of 80 mmHg. The reference pressure in the false lumen was assumed to be the same as in the true lumen. However, this value may change as a sensitivity study was performed on the false lumen pressure in this patient-specific case test, which is detailed in the results section. Moreover, it should be mentioned that the pure sliding boundary condition was applied on the cross-section at the two extremities and the outer surface of the adventitia was let free. The same material properties and simulation parameters as summarized in Table 2. Finally, it is worth noting that the non-uniform prestretches were used in this patient-specific artery, which was computed based on an iterative method previously developed by Laubrie et al. [25].

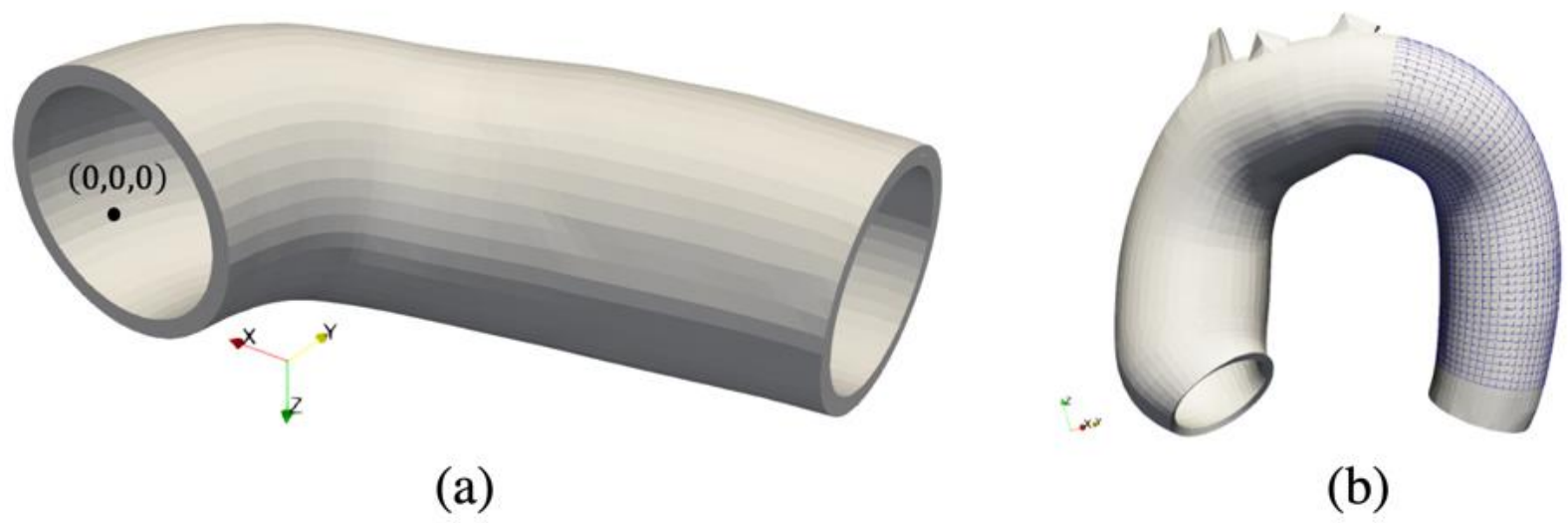

Figure 4 - (a) Schematic representation of the patient-specific human descending thoracic aortic segment on which the presented dissection model was applied for modeling G&R after cTBAD. (b) Illustration of the modeled aortic segment location, i.e., regions covered by the blue mesh, with respect to the whole patient-specific aortic arch.

In order to better describe the initial dissection tear of cTBAD, we introduce a numerical parameter, $\alpha$, for each circumferential cross-section of the arterial segment, as shown in Figure 5. Note that the unit vector normal to this cross-section is computed from the arterial centerline points. The unit vertical direction of the cross-section is approximately defined as the averaged projection vector of this section on the xy plane. After the definition of directions, we define $\alpha$ for each interfacial connecting spring

$$\alpha = \frac{l}{D} \tag{27}$$

where $l$ is the length of averaged spring positions, projected on the cross-section's vertical direction. The tear opening criterion is thus defined as $\alpha \geq \alpha_{min}$. Therefore, with $\alpha_{min}$ = 0, all interfacial springs between the adventitia and media layers will be broken, creating a full separation of the two layers. While if $\alpha_{min}$ is equal to 1, no interfacial springs will be broken and thus no presence of the tear. In this patient-specific simulation, the effect of the tear opening length to the G&R after cTBAD was considered, by varying the values of $\alpha_{min}$ from 0.8 (a narrow tear) to 0.5 (a wide tear).

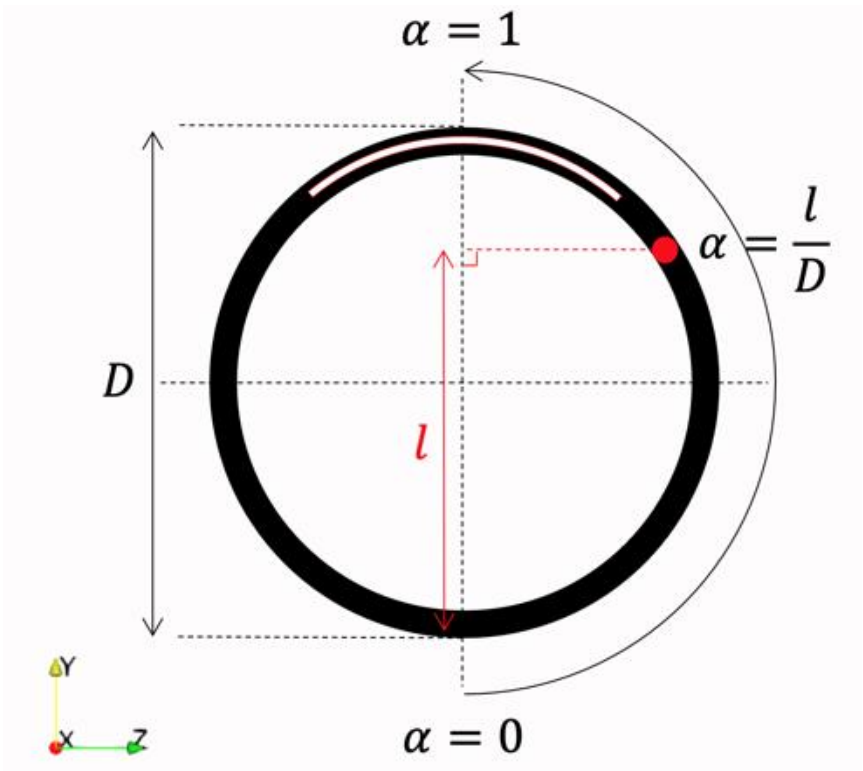

Figure 5 - Description of the initial tear opening criteria defined on each circumferential cross-section of the aortic segment.

### E. Computational details

All simulations were performed on a Macbook Pro with Intel Core i5 and 8 Go of memory. The computation time for each simulation takes around 2 hours. The low computation resources prove the computational efficiency of our dissection model.

## IV. Results

### A. Validation test case

Results of the validation test case are shown in Figure 6, illustrating the aneurysmal expansion of the arterial wall due to elastin loss. The evolution of the maximum inner radius of the aneurysm is shown, in comparison with the reference results [13], over a period of 10 years. The results indicated that the current two-continuum arterial wall model is in good agreement with the conventional single continuum arterial wall model. The aneurysmal expansion tends to recover its stability with a large gain parameter while a small gain parameter promotes an uncontrolled expansion of the aneurysm. With this validation test case, we justified the use of such a two-continuum arterial wall concept for G&R problems.

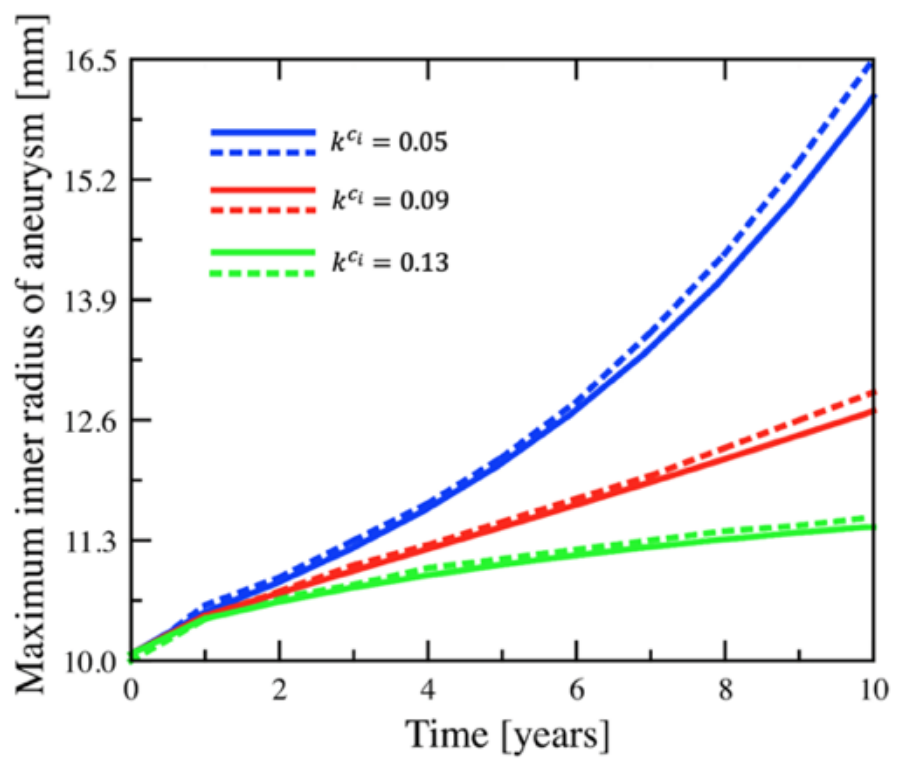

Figure 6 - Evolution of the maximum arterial inner radius over 10 years in response to an initial sudden elastin loss for both the two-continuum arterial wall model (solid lines) and the reference single-continuum arterial wall model (dash lines), with three different values of gain parameter related to collagen G&R [13].

### B. Application to a cylindrical artery

We first show the reference simulation results in the case of a cylindrical artery, as illustrated in Figure 7, with respect to a reference value of gain parameter $k_{\sigma}^{c_j} = k_{\sigma}^{m} = 0.05$. It can be seen that the dissected part of the artery, especially the outer adventitia layer, continues to dilate due to the effect of G&R after the initial tear opening. This aneurysmal dilatation tends to be unstable, with an increasing growth rate over time. Besides, the maximum stress, which is located at the vicinity of the tear edge, also increases rapidly over time.

Previous studies on G&R, which modeled aneurysm progression but disregarded effects of the dissection, reported that gain parameters have a determinant effect on the stability of aneurysmal dilatation[15], [16]. To investigate the effect of this gain parameter in the specific context of cTBAD, we considered three different values of gain parameters, ranging between 0.05 and 0.15, with results illustrated in Figure 8, showing the temporal evolution of the maximum outer diameter of the dissected cylindrical artery. It can be seen that the same tendency as previously reported in the literature has been observed for G&R after cTBAD. A small gain parameter tends to induce an unstable aneurysmal degeneration while a large gain parameter tends to favor the stability of aneurysmal dilatation.

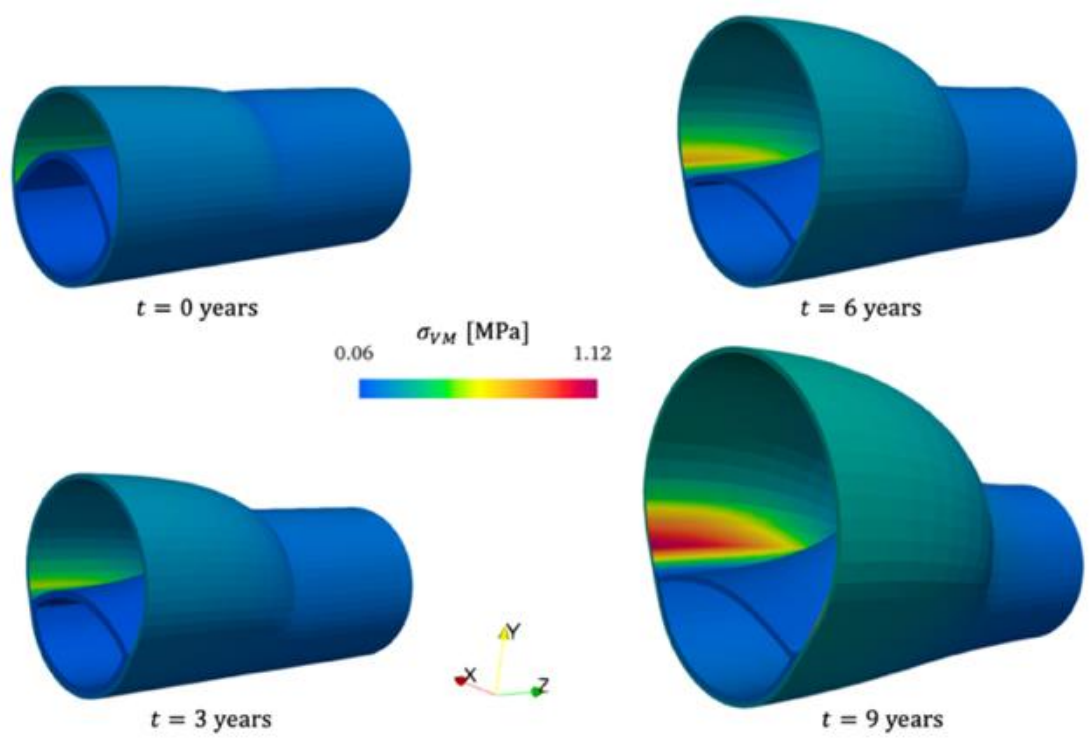

Figure 7 - Reference results of the cylindrical artery with respect to a reference value of gain parameter $k_\sigma^{cj} = k_\sigma^m = 0.05$, showing geometrical and equivalent von Mises stress evolutions after cTBAD over 9 years.

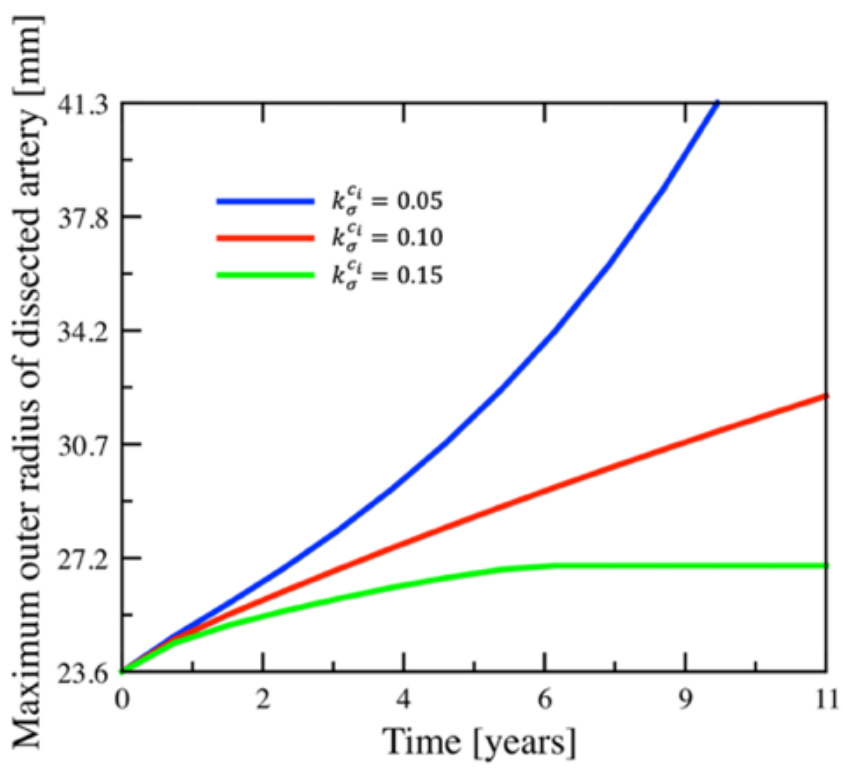

Figure 8 - Influence of the gain parameter to aneurysmal dilatation after cTBAD, showing the temporal evolution of the maximum outer diameter of the dissected cylindrical artery.

### C. Application to a toric artery

Reference simulation results in the case of a toric artery, with a reference value of gain parameter $k_\sigma^{cj} = k_\sigma^m = 0.05$, are shown in Figure 9. Similarly, the dissected part of the artery undergoes an unstable aneurysmal dilatation over 6 years after the initial tear opening. The maximum stress also increases over time. Besides, it is interesting to note that for this toric artery case, the stress seems to increase over the whole dissected adventitia layer, and is not limited to the vicinity of the tear edge as observed in the previous cylindrical artery case.

The effect of the gain parameter has also been investigated in this dissected toric artery. The temporal evolution of the maximum outer diameter of the dissected toric artery is shown in Figure 10. It can be seen that the results obtained are also in agreement with previous findings of the gain parameter: a large gain parameter favors a stable growth of aneurysm while a small gain parameter promotes an excessive enlargement of aneurysm after cTBAD.

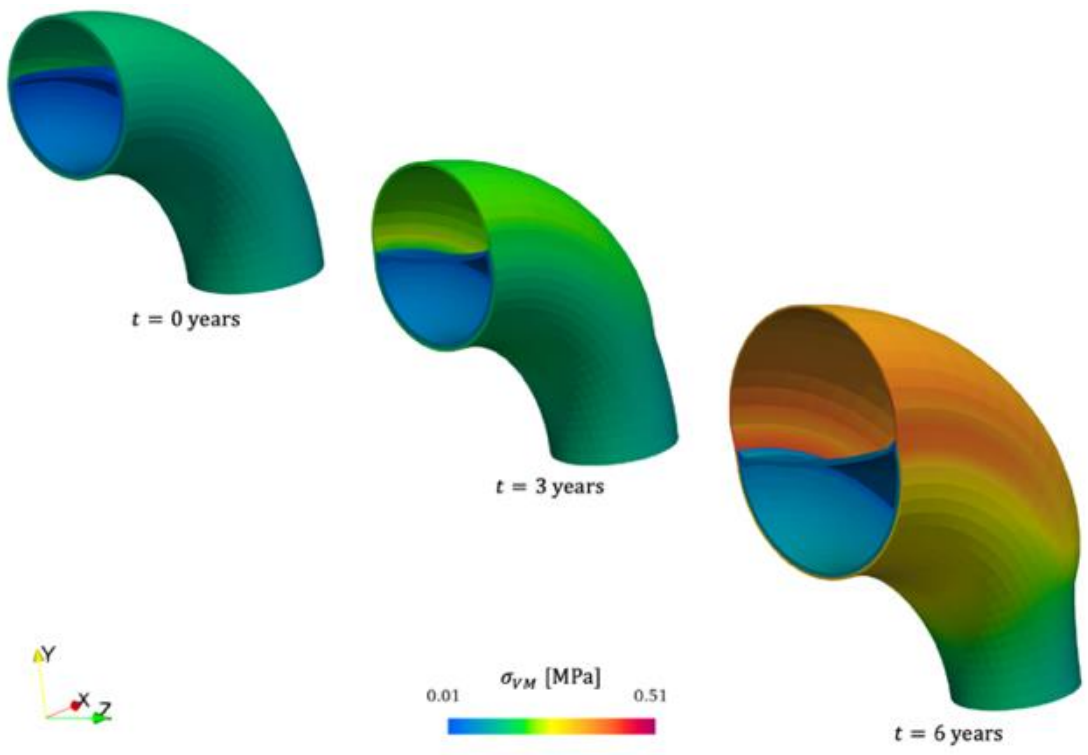

Figure 9 - Reference results of the toric artery with respect to a reference value of gain parameter $k_{\sigma}^{cj} = k_{\sigma}^{m} = 0.05$, showing geometrical and equivalent von Mises stress evolutions after cTBAD over 6 years.

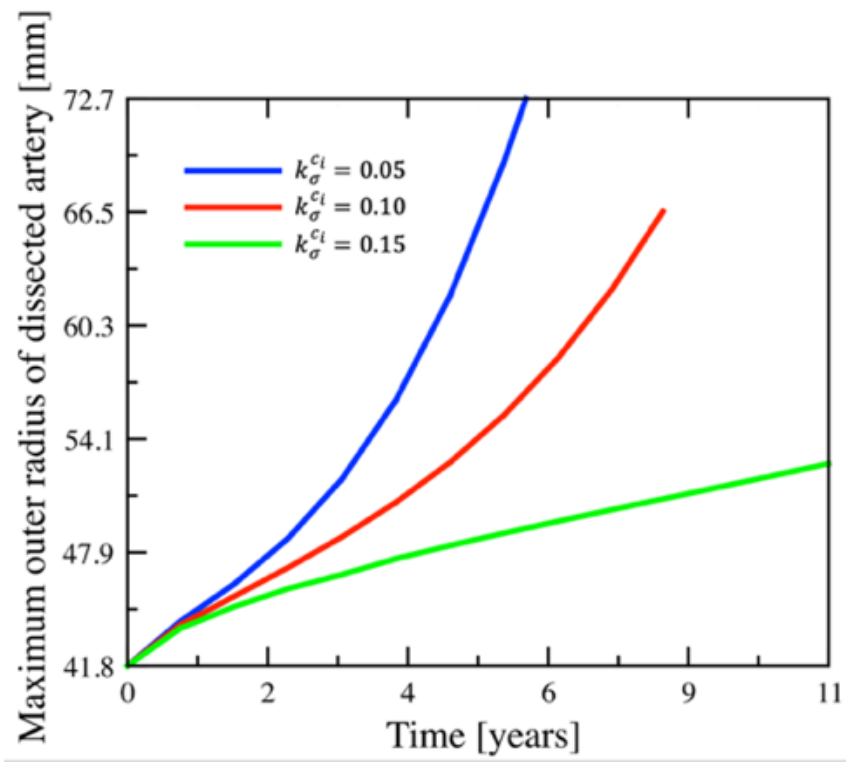

Figure 10 - Influence of the gain parameter to aneurysmal dilatation after cTBAD, showing the temporal evolution of the maximum outer diameter of the dissected toric artery.

## D. Application to a patient-specific artery

The reference simulation results in the case of a patient-specific artery, more precisely, a patient-specific human descending aortic segment, are shown in Figure 11, with respect to a reference value of gain parameter $k_{\sigma}^{cj} = k_{\sigma}^{m} = 0.05$. The results illustrate the geometrical and stress evolutions of the aortic segment, and also the circumferential cross-section at its dissected extremity. We found that the dissected aortic segment undergoes continuous aneurysmal dilatation overtime after the initial tear opening. Besides, it can be seen that there is a significant increase of stress over time, mostly on the dissected part of the outer adventitia layer.

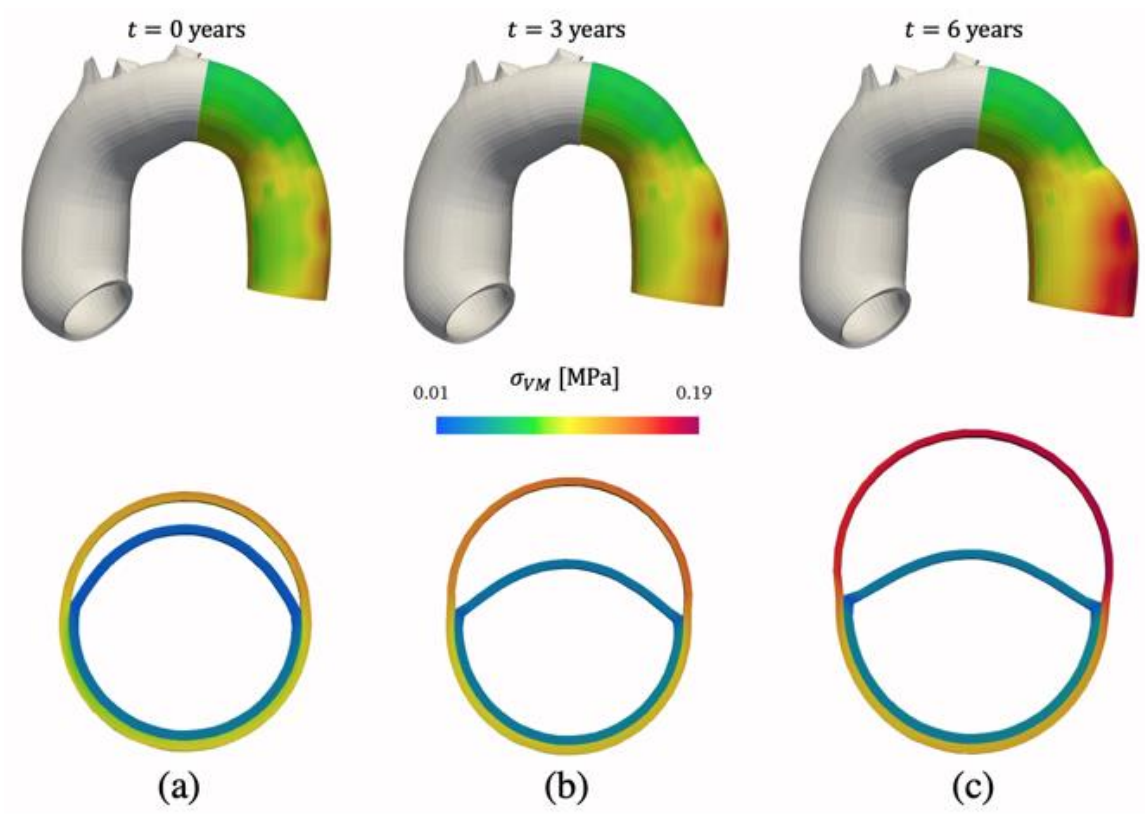

Figure 11 - Reference results of the patient-specific artery with respect to a reference value of gain parameter $k_\sigma^{cj} = k_\sigma^m = 0.05$, showing geometrical and equivalent von Mises stress evolutions after cTBAD over 6 years, as well as the dilated circumferential cross-section at its dissected maximum extremity, respectively at (a) 0 year of G&R, (b) 3 year of G&R and 6 year of G&R.

Similarly, the effect of the gain parameter on aneurysmal dilatation after cTBAD was investigated. Results obtained confirm the same tendency as observed in previous test cases: a large gain parameter tends to stabilize the aneurysmal dilatation of the dissected artery while a small gain parameter promotes an excessive aneurysmal degeneration after cTBAD.

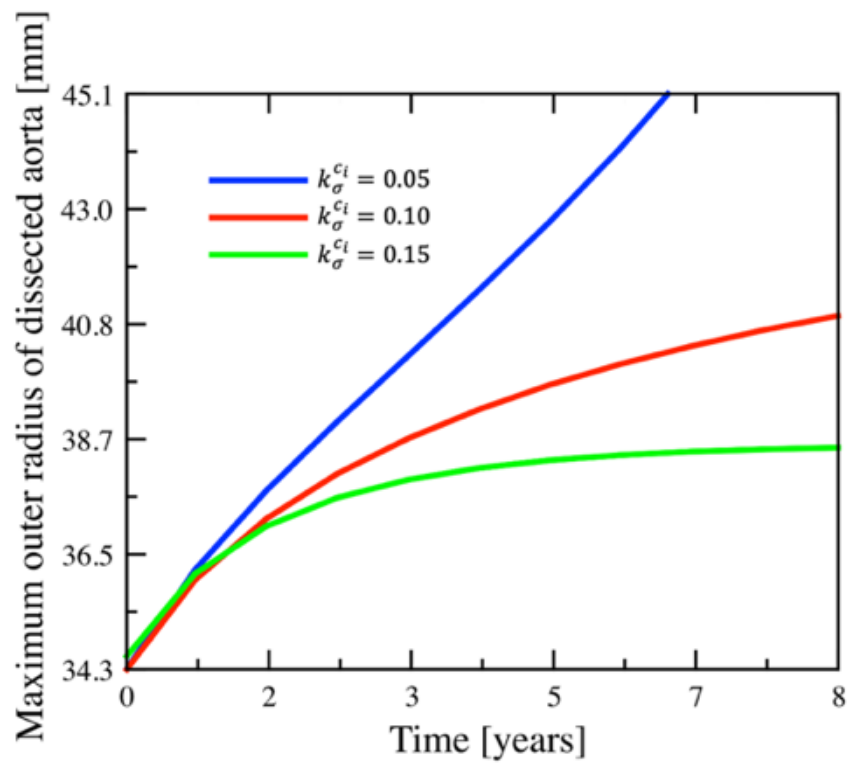

Figure 12 - Influence of the gain parameter to the aneurysmal dilatation after cTBAD, showing the temporal evolution of the maximum outer diameter of the dissected patient-specific artery.

Apart from the gain parameter, it has also been reported in the literature that the tear size may have a significant influence on aneurysmal dilatation after cTBAD [9], [26]. Being aware that effects of the tear size is very complex, which depend not only on the position of the tear but also on its irregular shape, involving generally measures in three dimensions. In this paper, note that we try to study only the cicumferential length of the tear, with four different values of $\alpha_{min}$ ranging from 0.8 to 0.5, i.e., from narrow circumferential tear to wide circumferential tear. Results obtained are shown in Figure 13, describing the temporal evolution of the maximum outer diameter of the dissected artery. It can be seen that a wider tear promotes an uncontrolled aneurysmal dilatation after cTBAD, while a narrow tear tends to favor a stable aneurysmal dilatation with a moderate growth rate.

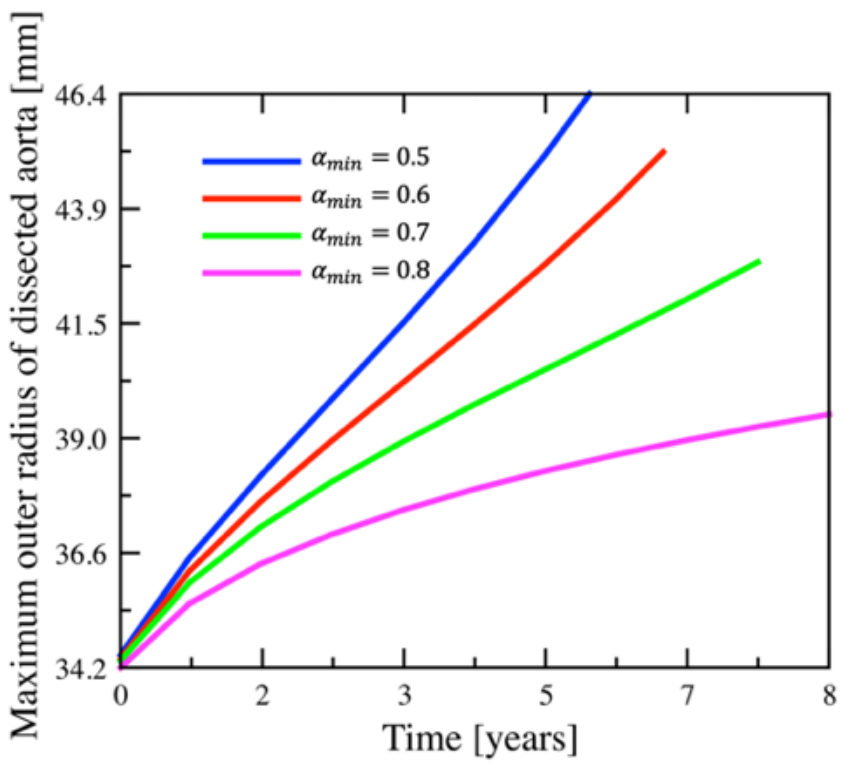

Figure 13 - Influence of the initial circumferential opening length of dissecting tear to the aneurysmal dilatation after cTBAD, showing the temporal evolution of the maximum outer diameter of the dissected patient-specific artery.

Finally, in this patient-specific simulation, we also investigated the effect of the pressure as it has been identified as a high-risk factor in cTBAD, especially in the false lumen where the pressure may impact directly on the stress distribution of the most weakened outer adventitia layer [6], [27], [28]. In order to evaluate the effect of the pressure in the false lumen, three different values of false lumen pressure are considered, respectively higher than the true lumen pressure ($p_{FL}/p_{TL} > 1$), equal to the true lumen pressure ($p_{FL}/p_{TL} = 1$), and lower than the true lumen pressure ($p_{FL}/p_{TL} < 1$). Note that pressure in the true lumen is assumed to be constant. Results are shown in Figure 14, showing the temporal evolution of the maximum outer diameter of the dissected artery. It can be seen that a higher pressure promotes aneurysmal dilatation after cTBAD. More precisely, regarding the aneurysmal dilatation rate at 3 years, a 10% pressure increase in the false lumen induces an "enlarging" growth of the dissected artery (≥ 3mm/year) while for false lumen pressure equal or below the true lumen pressure, aneurysmal dilatation can be considered as "stable" (< 3mm/year) [1].

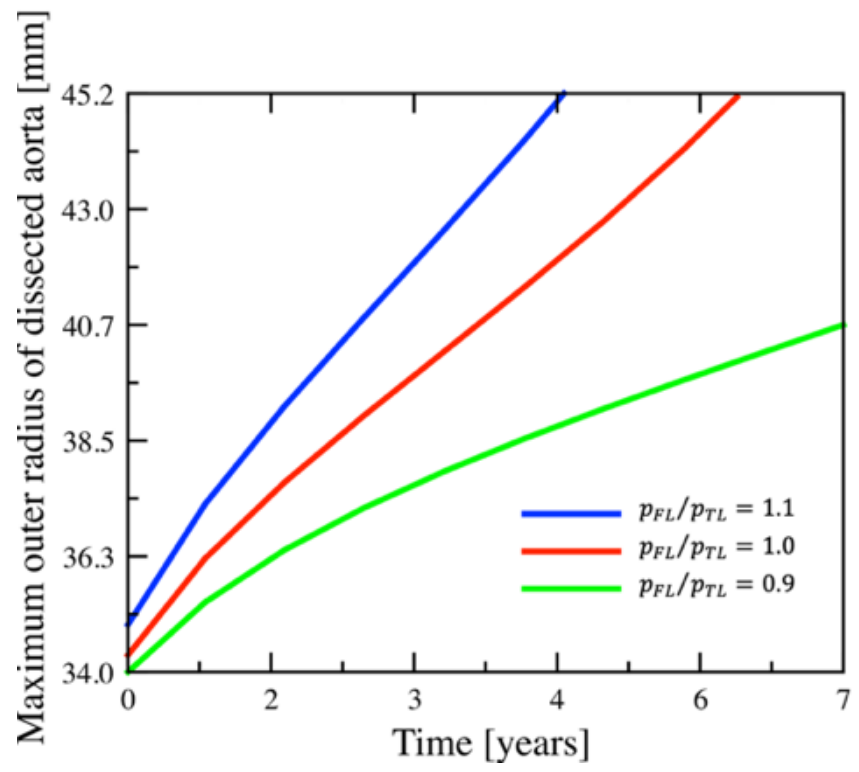

Figure 14 - Influence of the pressure in the false lumen to the aneurysmal dilatation after cTBAD, showing the temporal evolution of the maximum outer diameter of the dissected patient-specific artery.

# V. Discussion

cTBAD is associated with poor long term outcomes, mainly as a result of excessive aneurysmal dilatations. By consequence, a considerable part of patients with cTBAD will require ultimately surgical interventions such as endovascular repair or open surgery [29], [30]. However, there is a serious lack of risk assessment tools because our current understanding of the aneurysmal dilatation mechanism after cTBAD remains weak.

In this chapter, we proposed a numerical approach to study the role of G&R in aneurysmal dilatation in cTBAD. We found that the G&R process triggers naturally the aneurysmal dilatation. Moreover, it was found that with a large gain parameter related to collagen G&R, the aneurysmal dilatation tends to be stable while with a small gain parameter, there would be an excessive aneurysmal degeneration. It is interesting to note that the results obtained are in agreement with clinical evidence reported by Juvonen et al. [6], where older patients present a higher risk of aneurysmal rupture in case of cTBAD. In fact, this gain parameter describes the capacity of arteries to restore its tensional equilibrium state in case of a disturbance of its mechanobiological equilibrium [14] and it has been reported that age may affect this gain parameter with older patients generally having an impaired stress-regulated repair mechanism compared to young patients [31].

Based on sensitivity analysis performed on patient-specific simulations, we found that the circumferential tear length also has a significant influence on the G&R process after cTBAD. A wide tear promotes an unstable development of aneurysmal dilatation while a narrow tear reduces the risk of uncontrolled aneurysmal dilatation. The reason could be twofold. First, a wide initial opening tear means naturally a larger initial dissected arterial diameter once the false lumen is pressurized compared to a narrow initial tear. Secondly, the consequence of a larger initial dissected arterial diameter is that the deformation and stress will be much higher, especially in the dissected outer adventitia layer, accelerating the G&R process of the artery. Indeed, the results remained very limited, considering the irregular three-dimensional tear shape and other complex mechanobiological phenomena neglected. However, it provides a mechanical proof that the tear size is also an important influencing factor that needs to be considered in the risk assessment of patients with cTBAD.

Finally, our results indicated that the pressure in the false lumen has a determinant role in the aneurysmal progression rate of the dissected artery. We found that a relative 10% increase of pressure in the false lumen, compared to that of the true lumen, is sufficient to promote an "unstable" growth of the dissecting aneurysm. This can be critical for patients, as surgical interventions are usually recommended for such situations [1].

Despite the above promising results obtained, the current model still has some shortcomings that could be addressed in the future. First, in the present model, the tear configuration remains fixed after its initial creation. However, with the continuous aneurysmal dilatation and accumulation of stress especially near the edge of the tear, the initial tear may propagate due to high-stress concentration and thus alter the G&R process. Therefore, integration of the tear propagation models, such as that reported in the literature [32]-[35], could be an essential step to build a more reliable dissection model for evaluating the G&R effect to aneurysmal dilatation after cTBAD.

Secondly, the dissection model is relatively simplified. Intraluminal thrombus, which was often reported to have an important role in dissecting aneurysm pathologies and rupture [36], [37], was literally neglected in this work. Besides, the effects of surrounding tissues have also been neglected although it has already been reported in the literature that the surrounding connective tissues or vertebral column may impact regional adaptation of aortic walls by changing the local wall stress distribution or even more directly the aneurysmal aortic shape [38], [39]. Moreover, potential dynamic effects of the blood flow inside the arterial wall are currently not taken into consideration, despite its non-negligible effect directly on the wall stress distribution, as reported in the literature [26], [40].

Finally, the initial tear configuration was restricted to its circumferential length. Location of the tear, number of the tears or other dimensions related to the tears are currently neglected. For further patient-specific simulations, these parameters should be carefully considered as they may directly impact the pressure in both the false lumen and the true lumen [9].

# VI. Conclusions

In summary, we introduced an efficient 3D finite-element model, based on an open-source in-house code, to model the aneurysmal dilatation due to G&R after cTBAD. We showed the potential of this dissection model to simulate G&R process after cTBAD, from simple test cases with idealized arterial geometries to a more relevant case with a patient-specific geometry. The effects of different parameters on aneurysmal dilatation were assessed through a comprehensive sensitivity analysis. It was found that the gain parameter related to collagen G&R as well as the circumferential initial tear length, has an undeniable impact on the stability of the dissecting aneurysm. Moreover, our results indicated that the stability of the dissecting aneurysm is very sensitive to the intraluminal false lumen pressure. A relative pressure increase of 10% in the false lumen may induce an excessive aneurysmal degeneration in patients with cTBAD.

Future work is twofold. The first one is coupling the present dissection model with tear propagation models, for applications to more reliable patient-specific simulations. The second one is to account for a more accurate configuration of opening tears while considering the potential dynamic effects of the blood flow inside the dissected arteries.